\documentstyle[preprint,prl,aps]{revtex}

\begin{document}

% uncomment \draft to have PACS numbers appear
\draft

% put preprint numbers.
\preprint{\vbox{\hfill OSU--1101\\ \null\hfill UTHEP--268}}

\narrowtext

%\noindent
%{\bf Comment on: ``Nucleon-nucleon scattering lengths
%             in QCD sum rules''}

\title{Comment on: ``Nucleon-nucleon scattering lengths\\
             in QCD sum rules''}

\author{R. J. Furnstahl}
\address{Department of Physics \\
         The Ohio State University,\ \ Columbus, Ohio\ \ 43210}
\author{T. Hatsuda}
\address{Institute of Physics \\
University of Tsukuba,\ \ Tsukuba, Ibaraki 305, Japan}
\date{December, 1993}
\maketitle

\bigskip
\bigskip

In a recent Letter \cite{KM2}, Kondo and Morimatsu present a QCD
sum rule calculation of nucleon-nucleon scattering lengths.
They also
relate the empirical scattering lengths to
the nucleon mass shift $\delta M$ in nuclear matter to cast doubt on the
``linear density approximation.''
In this Comment, we point out  flaws in both parts of
their analysis.

It is useful to start with the (nonrelativistic)
optical potential for a nucleon in nuclear matter in the
lowest order of the multiple scattering expansion \cite{EIS}
\begin{equation}
	V({\bf q}) \sim \int^{k_{{\scriptscriptstyle\rm F}}}
        \! \tau ({\bf q},{\bf k})\, d^3k\,
                 \ ,
	\label{eq:shift}
\end{equation}
where $\tau({\bf q},{\bf k})$ is the effective forward-scattering amplitude
 for an  incoming nucleon with momentum ${\bf q}$ and a nucleon in nuclear
matter with momentum ${\bf k}$ \cite{EIS}.

If $|{\bf q}| \gg k_{{\scriptscriptstyle\rm F}}$,  one can
 approximate $V({\bf q})$ by neglecting the momenta of nucleons
in the Fermi sea and replacing $\tau({\bf q},{\bf k})$ by the
 amplitude in  free space $t({\bf q},{\bf k})$:
$	V({\bf q}) \sim t ({\bf q},{\bf 0})\, \rho\,$,
which is the usual impulse approximation.
However, in Ref.~\cite{KM2}, ${\bf q}=0$ is taken {\it after\/}
the impulse approximation.
 Applying this approximation with $|{\bf q}| \ll k_F$ drastically
 overestimates the mass shift.
In \cite{KM2}, the scattering lengths are used to predict
 ${\rm Re\,} V({\bf 0}) \simeq -600$ MeV at saturation density.
 ($\delta M$  in Ref.~\cite{KM2}
is $-{\rm Re\,} V({\bf 0})$ here.)

If formula (\ref{eq:shift}) is used with ${\bf q}=0$, one should conclude
instead that
$	V({\bf 0}) \sim \int^{k_{{\scriptscriptstyle\rm F}}}
                   \tau({\bf 0},{\bf k}) d^3k\, $.
 Then it is clear that  $t({\bf 0},{\bf 0})$,
which is determined by the large scattering lengths
 due to the (near) zero-energy bound states in the s-wave channel,
has little to do with $\delta M$ at nuclear matter density.
Instead, the main effect is the average interaction of the zero-momentum
``incoming'' nucleon with the finite momentum nucleons in the Fermi
sea.
 The antisymmetrization of the projectile
  with the target nucleons  neglected in (\ref{eq:shift}) also
 becomes  important for small ${\bf q}$;    detailed calculations
 show $|{\rm Re\,} V({\bf 0})| < 100$ MeV \cite{EIS}.

To obtain a meaningful expression for $\delta M$ that is
linear in the density,
one can pull out the average of
the effective nucleon-nucleon interaction in medium.
Thus, the ``linear density approximation'' follows naturally
from a mean-field
treatment of the nucleon optical potential or self-energy in medium.
%rather than from a derivative with respect to density.
This is compatible with finite-density QCD sum rules, since an
extrapolation from short times implies that mean-field physics should be
best reproduced, and the leading condensates are apparently
well-approximated as being linear in density \cite{lp1}.
 For this reason,
the sum rule calculations in Ref.~\cite{lp1} focus separately on
the  Lorentz scalar and
vector self-energies, because relativistic phenomenology
suggests that the mean-field components dominate.
In contrast,
$\delta M$ involves
large cancellations between these components,
which imply that neglected effects may become important.

Despite all this,
the QCD sum rule calculation presented in Ref.~\cite{KM2}
seems to predict qualitatively correct
scattering lengths.
We argue that this does not make sense physically;
if one applies sum rule methods
to calculate scattering lengths, one should not expect
to recover details about (near) zero-energy bound states.
{\it In fact, the sum rule calculation in Ref.~\cite{KM2}
is not correct, and also greatly  overestimates
the mass shift}.

The sum rule calculation in Ref.~\cite{KM2}
 follows from the finite-density sum rules of Ref.~\cite{KM1} by taking
a derivative with respect to the density.
However,  in Ref.~\cite{FURNSTAHL93}, it is shown that the sum rules in
Ref.~\cite{KM1}  are
missing an important element, {\it i.e.},  part  of the
continuum contribution, which is asymmetric at finite density.
In some formulations \cite{lp1,HATSUDA91}
this is numerically unimportant, but in the present case,
contributions from the asymmetric continuum are large and
completely alter the results \cite{FURNSTAHL93}.
As a result, conclusions drawn in Ref.~\cite{KM2}
from this sum rule about the size of the implied
mass shift and the role of the vector density of quarks are incorrect.

In summary,
our conclusions are opposite to those of Ref.~\cite{KM2}.
In a consistent QCD sum rule calculation,
nucleon mass shifts in nuclear matter are not overestimated by
sum rules that approximate condensates to linear order in the
density.
Furthermore, {\it empirical\/}
nucleon-nucleon scattering lengths have little
to do with these mass shifts.
Finally, the sum rule calculations of Refs.~\cite{KM2,KM1},
when corrected, give results
consistent with Ref.~\cite{lp1}, but
are much more sensitive to details.



\begin{references}
%
\bibitem{KM2}Y. Kondo and O. Morimatsu, Phys.\ Rev.\ Lett. {\bf 71},
     2855 (1993).
%
\bibitem{EIS} H. Feshbach, {\it Theoretical Nuclear Physics:
 Nuclear Reactions}, (John Wiley \& Sons, New York, 1992).
%
\bibitem{lp1}R.~J. Furnstahl, D.~K. Griegel, and T.~D. Cohen,
        Phys.\ Rev.\ C {\bf 46}, 1507 (1992).
%
\bibitem{KM1}Y. Kondo and O. Morimatsu, Institute for Nuclear Study report
   INS-Rep.-933, (June 1992).
%
\bibitem{FURNSTAHL93}R. J. Furnstahl, Ohio State University Report
        \#OSU--0901, nucl-th/9311002, (1993).
%
\bibitem{HATSUDA91}T. Hatsuda, H. H{\o}gaasen, and M. Prakash,
        Phys.\ Rev.\ Lett.\ {\bf 66}, 2851 (1991).


%
\end{references}
\end{document}